\documentclass[12pt,twoside]{article}

\usepackage[utf8]{inputenc} 
\DeclareUnicodeCharacter{2010}{-}
\usepackage{placeins} 
\usepackage[colorlinks,bookmarks=false,linkcolor=blue,urlcolor=blue]{hyperref}
 \usepackage{authblk} 
\usepackage{fancyhdr}
\usepackage[toc,page]{appendix} 
\usepackage{soul} 
\usepackage{caption}
\usepackage{subcaption}
\usepackage{amsmath}
\usepackage{amssymb}
\usepackage{amsfonts}
\usepackage{siunitx}
\usepackage{mathtools} 
\usepackage{xcolor} 
\usepackage{tikz}
\usetikzlibrary{arrows,shapes}
\usepackage{bm}
\usepackage{bbold} 
\usepackage{listings}
\definecolor{dkgreen}{rgb}{0,0.6,0}
\definecolor{gray}{rgb}{0.5,0.5,0.5}
\definecolor{mauve}{rgb}{0.58,0,0.82}
\newcommand{\tw}[1]{\texttt{#1}}
\def \bes {\begin{equation*}}
\def \be {\begin{equation}}
\def \ees {\end{equation*}}
\def \ee {\end{equation}}
\def \dd  {{\rm d}}

\newcommand{\bigO}[1]{O(#1)}

\newcommand{\nrm}[1]{\left\lvert#1\right\rvert}

\newcommand{\rms}[1]{\left[ #1 \right]_{\mathrm{rms}}}
\newcommand{\deltaf}{\delta f}

\newcommand{\eqf}{F_{0,\spec}}
\newcommand{\phiturb}{\varphi}

\newcommand{\phiturbz}{\varphi_{\mathrm{Z}}}
\newcommand{\phiturbnz}{\varphi_{\mathrm{NZ}}}

\newcommand{\Znum}{Z}
\newcommand{\charge}{e\Znum}

\newcommand{\rhostarspec}{\rho_{*\spec}}

\newcommand{\gslength}{a}

\newcommand{\torfreq}{\Omega_\torang}

\newcommand{\gammaE}{\gamma_E}
\newcommand{\gammaz}{\gamma_{\mathrm{Z}}}

\newcommand{\Tfloq}{T_{\mathrm{F}}}
\newcommand{\gammaavg}{\langle\gamma\rangle_t}
\newcommand{\gammamax}{\gamma_{\mathrm{max}}}
\newcommand{\taunl}{\tau}
\newcommand{\corrfunc}{\mathrm{Cor}}
\newcommand{\lcorrx}{\ell_x}
\newcommand{\lcorrxzonal}{\ell_{x,\mathrm{Z}}}
\newcommand{\lcorry}{\ell_y}
\newcommand{\deltax}{\delta_x}
\newcommand{\deltay}{\delta_y}
\newcommand{\energy}{\varepsilon_{\spec}}
\newcommand{\magmom}{\mu_{\spec}}
\newcommand{\relvel}{w}

\newcommand{\uflow}{u}

\newcommand{\vbs}{\bm{V}_{B,\spec}}

\newcommand{\vcs}{\bm{V}_{C,\spec}}
\newcommand{\veturb}{\bm{V}_E}
\newcommand{\vthi}{v_{\mathrm{th},i}} 
\newcommand{\vzonal}{V_{\mathrm{Z}}}
\newcommand{\bunit}{\hat{\bm{b}}}

\newcommand{\shat}{\hat{s}}
\newcommand{\torang}{\phi}
\newcommand{\polang}{\theta}

\newcommand{\rmaj}{R}
\newcommand{\gradius}{\rho}
\newcommand{\gang}{\vartheta}
\newcommand{\gcenter}{\bm{R}_\spec}
\newcommand{\ppos}{\bm{r}}
\newcommand{\rpsi}{r_\psi}
\newcommand{\rpsimid}{r_{\psi,\midline}}

\newcommand{\magflux}{\psi}
\newcommand{\kxshear}{k_x}

\newcommand{\Kx}{K_x}
\newcommand{\Ky}{K_y}
\newcommand{\Deltakx}{\Delta\kxshear}
\newcommand{\Deltaky}{\Delta k_y}

\newcommand{\spec}{s}

\newcommand{\midline}{0}
\newcommand{\rfr}{r}
\newcommand{\Rmaj}{R_\psi}
\newcommand{\elong}{\kappa}
\newcommand{\triang}{\delta}

\newcommand{\gavg}[1]{\langle #1 \rangle_{\gcenter}}
\newcommand{\ringavg}[1]{\langle #1 \rangle_{\ppos}}
\newcommand{\flxsurfavg}[1]{\langle #1 \rangle_{\magflux}}
\newcommand{\flxsurfavgbig}[1]{\Big\langle #1 \Big\rangle_{\magflux}}

\newcommand{\LTs}{L_{T_\spec}}
\newcommand{\LTi}{L_{T_i}}
\newcommand{\LTe}{L_{T_e}}

\newcommand{\Lni}{L_{n_i}}
\newcommand{\Lne}{L_{n_e}}

\newcommand{\dsub}[2]{%
  \setsepchar{_}%
  \readlist\mymat{#1}%
  \ifnum\mymatlen>1{\mymat[1]_{\mymat[2],#2}}%
  \else{{#1}_#2}\fi%
}
\newcommand{\Qgbohm}{Q_{\mathrm{gB}}}

\newcommand{\boltzcst}{k_{B}}
\usepackage[utf8]{inputenc} 
\DeclareUnicodeCharacter{2010}{-}
\usepackage{placeins} 
\usepackage[colorlinks,bookmarks=false,linkcolor=blue,urlcolor=blue]{hyperref}
 \usepackage{authblk} 
\usepackage{fancyhdr}
\usepackage[toc,page]{appendix} 
\usepackage{soul} 
\usepackage{caption}
\usepackage{subcaption}
\usepackage{amsmath}
\usepackage{amssymb}
\usepackage{amsfonts}
\usepackage{siunitx}
\usepackage{mathtools} 
\usepackage{xcolor} 
\usepackage{tikz}
\usetikzlibrary{arrows,shapes}
\usepackage{bm}
\usepackage{bbold} 
\usepackage{listings}
\definecolor{dkgreen}{rgb}{0,0.6,0}
\definecolor{gray}{rgb}{0.5,0.5,0.5}
\definecolor{mauve}{rgb}{0.58,0,0.82}
\begin{document}

\title{Bistable turbulence in strongly magnetised plasmas with a sheared mean flow}
\date{\small\today}
\author[1,2]{N. Christen \thanks{Contact: nicolas.christen@physics.ox.ac.uk}}
\author[1,3]{M. Barnes}
\author[1]{M. R. Hardman}
\author[1,4]{A. A. Schekochihin}
\affil[1]{Rudolf Peierls Centre for Theoretical Physics, University of Oxford, Oxford OX1 3PU, UK}
\affil[2]{Lincoln College, Oxford OX1 3DR, UK}
\affil[3]{University College, Oxford OX1 4BH, UK}
\affil[4]{Merton College, Oxford OX1 4JD, UK}
\maketitle

\begin{abstract}

The prevailing paradigm for plasma turbulence associates a unique stationary state to given equilibrium parameters. We report the discovery of bistable turbulence in a strongly magnetised plasma with a sheared mean flow. Two distinct states, obtained with identical equilibrium parameters in first-principle gyrokinetic simulations, have turbulent fluxes of particles, momentum and energy that differ by an order of magnitude -- with the low-transport state agreeing with experimental observations. Occurrences of the two states are regulated by the competition between an externally imposed mean flow shear and ``zonal'' flows generated by the plasma. With small turbulent amplitudes, zonal flows have little impact, and the mean shear causes turbulence to saturate in a low-transport state. With larger amplitudes, the zonal shear can (partially) oppose the effect of the mean shear, allowing the system to sustain a high-transport state. This poses a new challenge for research that has so far assumed a uniquely defined turbulent state.

\end{abstract}


\section{Introduction} \label{sec:introduction}

Turbulence is a common feature of magnetised plasmas, appearing in systems as varied as the solar wind, astrophysical accretion disks and laboratory plasmas. According to the most common paradigm for such systems, a unique stationary turbulent state can be identified given a certain stirring mechanism and a set of equilibrium plasma parameters. Multistable solutions -- for which identical parameters admit distinct turbulent states -- are known to occur in neutral fluids \cite{snedekerAIAA66, burggrafJFM77, schmuckerFDR88}, where they are associated with bifurcations and hysteretic behaviour \cite{shternPoF93, raveletPRL04}. Multistability has also been reported in weakly magnetised systems of charged fluids \cite{simitevEPL09, latterRAS12}. In this work, we report the discovery of bistable turbulence in a strongly magnetised plasma, using direct numerical simulations. We find that bistability arises in such a plasma through the interplay of two crucial mechanisms: an externally imposed mean flow shear and self-generated ``zonal'' flows. Our observations are made in a toroidal geometry typically encountered in magnetic-confinement-fusion experiments, though the results may be generalisable to other systems.

Previous studies have already established that sheared flows play an important role in regulating turbulence. In the absence of an externally imposed mean flow shear, the plasma is known to generate sheared ``zonal'' flows spontaneously, contributing to the saturation of turbulence \cite{biglariPoF90, dimitsPoP00, rogersPRL00, diamondPPCF05, colyerPPCF17, vanwykPPCF17, ivanovJPP20}. So far, it is therefore usually assumed that the effect of zonal flows is to suppress turbulence. When a mean flow shear is imposed, it provides an additional mechanism for suppressing turbulence. Specifically, the shear in the mean flow perpendicular to the magnetic field has been found to reduce turbulent fluctuations \cite{vanwykPPCF17, waelbroeckPoF91, artunPoF92, dimitsPRL96, synakowskiPRL97, manticaPRL09, cassonPoP09, highcockPRL10, barnesPRL11, highcockPRL12, schekochihinPPCF12, vanwykJPP16}. It has also been shown that the effect of the mean shear weakens away from marginal stability \cite{foxPPCF17, vanwykPPCF17}.

Here, we find that the transition from turbulent states where the mean shear plays an essential role to states where it appears to matter only marginally is characterised by a discontinuous jump in the level of turbulent transport. Most importantly, we show that in a region of parameter space near this transition, two distinct turbulent states exist with identical equilibrium parameters but dramatically different levels of transport. We find that the presence of strong zonal flows is a feature of higher-transport states -- the opposite of what is usually assumed. The main result is presented in figure \ref{fig:hysteresis}.

This discovery has important implications for research in nuclear fusion. In experiments, turbulent fluxes are set by the external injection of particles, heat and momentum into the plasma, with profile gradients evolving until a stationary state is reached. However, because of computational cost, direct numerical simulations only consider a small fraction of the device's volume, in which they solve the inverse problem: for given local equilibrium quantities (e.g., profile gradients), the simulations determine the associated turbulent fluxes. The flux-to-gradient problem and its inverse can be considered equivalent if a one-to-one correspondence exists between the turbulent transport and the equilibrium parameters. Our work shows that this correspondence is not always one-to-one, which poses a challenge for modelling transport -- and thus for designing future fusion devices. Finally, bistability has some remarkable consequences, such as the possibility for bifurcations of turbulent transport and gradient-relaxation cycles to develop (these have previously been considered in the absence of mean flow shear \cite{peetersPoP16}).


\section{Modelling plasmas with a sheared mean flow} \label{subsec:model}

We consider equilibrium parameters obtained from a fusion experiment conducted at the Joint European Torus facility (discharge \#68448 \cite{sirenJI19}). The plasma is confined by magnetic fields that trace out nested toroidal surfaces, with the equilibrium density and temperature staying constant along the field lines. External heating sources sustain an ion temperature gradient between the hotter core and the colder edge of the plasma, which then drives the dominant linear instability \cite{romanelliPoF89, cowleyPoF91}. A sheared mean toroidal flow is generated by injecting beams of neutral hydrogenic atoms into the plasma. The ratio of thermal to magnetic pressure is small, so the turbulent fluctuations can be assumed electrostatic. We also neglect any trace impurities in the plasma and only consider two kinetic species -- the electrons and the main deuterium ions. The simulations include collisions, as well as a small amount of numerical hyperviscosity \cite{belliPhD}. The numerical parameters used for this work are provided in the appendix.

The model used for this work is presented in appendix \ref{sec:gk_system}. The time evolution of turbulent fluctuations is described by following the local $\delta f$ gyrokinetic approximation \cite{cattoPP78, friemanPoF82, sugamaPoP98, abelRPP13}. This approximation relies on the scale separations present in the system by defining an asymptotic-expansion parameter $\rhostarspec = \gradius_\spec/\gslength \ll 1$ for a species $\spec$, where $\gradius_\spec$ is the particle's gyroradius around a magnetic field line and $\gslength$ is the minor radius of the device. As a result, the rapid gyromotion of particles can be averaged out. In this approach, the kinetic equation and the quasineutrality condition form a closed system of equations for the fluctuating probability distribution function of charged rings and for the electrostatic potential $\phiturb$. The system is solved numerically with the code \tw{GS2} \cite{kotschenreutherCPC95, barnesPoP09, highcockThesis12} in a filament-like simulation domain \cite{beerPoP95} that follows a magnetic field line as it wraps around the torus. The code then computes the turbulent contributions to the heat and momentum fluxes exiting the core of the plasma, which we denote by $Q_\spec$ and $\Pi_\spec$, respectively. In the following figures, we normalise $Q_\spec$ to the so-called gyro-Bohm value $\Qgbohm = \flxsurfavg{\nrm{\nabla\magflux}}n_iT_i\vthi\rhostarspec^2$, where $\magflux$ is the poloidal magnetic flux, $\flxsurfavg{\cdot}$ the average over a magnetic-flux surface, $T_i$ the ion temperature multiplied by the Boltzmann constant $\boltzcst$, $\vthi=\sqrt{2 T_i/m_i}$ the ion thermal speed and $m_i$ the ion mass. The ion temperature gradient is specified through $\gslength/\LTi$, where $\LTi$ is the local $e$-folding length scale of $T_i$. Finally we denote by $\gammaE$ the rate at which the mean flow is sheared across magnetic surfaces, normalised by $\vthi/\gslength$.

When $\gammaE\neq 0$, linear modes (known as Floquet modes) are advected along the magnetic field lines, passing through regions of the plasma that are alternately stable (inboard of the torus) and unstable (outboard) to the ion temperature gradient \cite{waelbroeckPoF91}. As a consequence, their linear growth rate is time dependent with a Floquet period $\Tfloq = 2\pi\shat/\gammaE$. Here, $\shat$ (defined in appendix \ref{sec:gk_system}) measures how the twisting of magnetic field lines around the torus changes with the minor radius. In the following, we denote the time-averaged growth rate by $\gammaavg$, and the maximum instantaneous growth rate by $\gammamax$. In this work, the code \tw{GS2} was used with an improved algorithm for background flow shear described in \cite{mcmillanPPCF19, christenJPP21}, although it has been verified that the same conclusions are reached when using the original algorithm devised by \cite{hammettAPS06}.

\section{Two distinct turbulent states}

\begin{figure}
\centering
\includegraphics[scale=0.6]{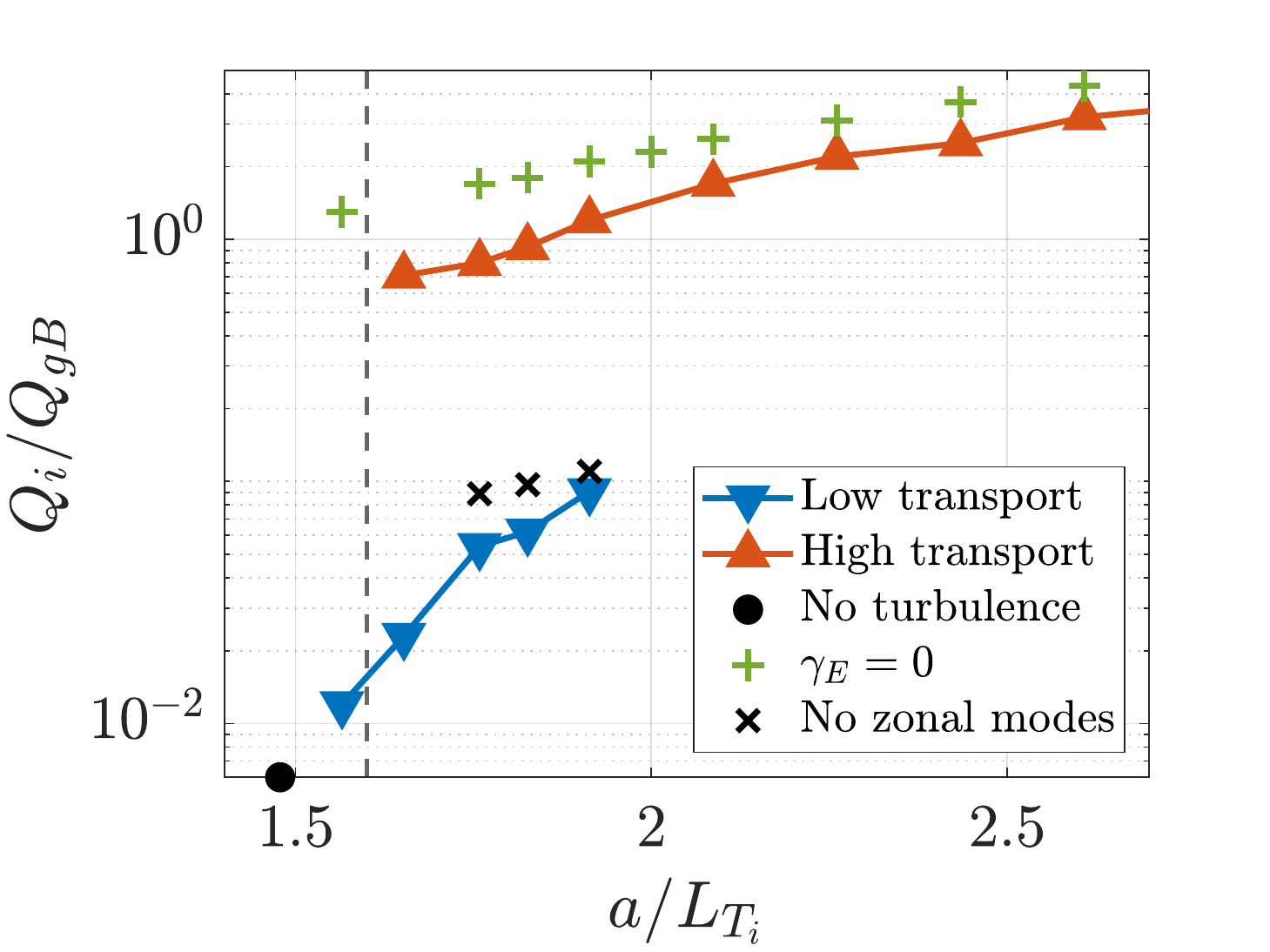}
\caption{Dependence of the turbulent ion heat flux on the inverse ion-temperature-gradient scale length. In the simulations labelled by green `+' signs, the externally imposed mean flow shear was set to zero. For all other simulations, $\gammaE=-0.079$. Zonal modes are artificially zeroed out in simulations labelled by black crosses. The black circle denotes a simulation where amplitudes decay with time and no saturated turbulent state is observed. The dashed line marks the temperature gradient below which there is no effective linear instability ($\gammaavg < 0$) in the presence of mean flow shear.}
\label{fig:hysteresis}
\end{figure}

Near marginal stability, we find that two distinct turbulent states can be obtained at identical equilibrium parameters. This is shown in figure \ref{fig:hysteresis} where saturated values of $Q_i/\Qgbohm$ are plotted against $\gslength/\LTi$ for a particular value of $\gammaE$. We find that the fluxes computed in the low-transport state match the levels of transport observed in the experiment, while the fluxes computed in the high-transport state differ from it by an order of magnitude. For equilibrium parameters where the two states exist, it is the initial size of the fluctuation amplitudes that determines which state is observed in a simulation. While the impact of initial conditions on gyrokinetic simulations was explored in previous work such as \cite{pueschelPoP08}, our work is the first to obtain two distinct, saturated and finite-amplitude turbulent states with identical equilibrium parameters. Interestingly, we note that both low-transport and high-transport states can exist above and below the threshold for linear instability. Previous work had already established this for a single, finite-amplitude turbulent state sustained either by a linear instability (known as supercritical turbulence) or by transient linear growth (known as subcritical turbulence \cite{highcockPRL10, barnesPRL11, schekochihinPPCF12}).

\FloatBarrier

\begin{figure}
\centering
\includegraphics[scale=0.38]{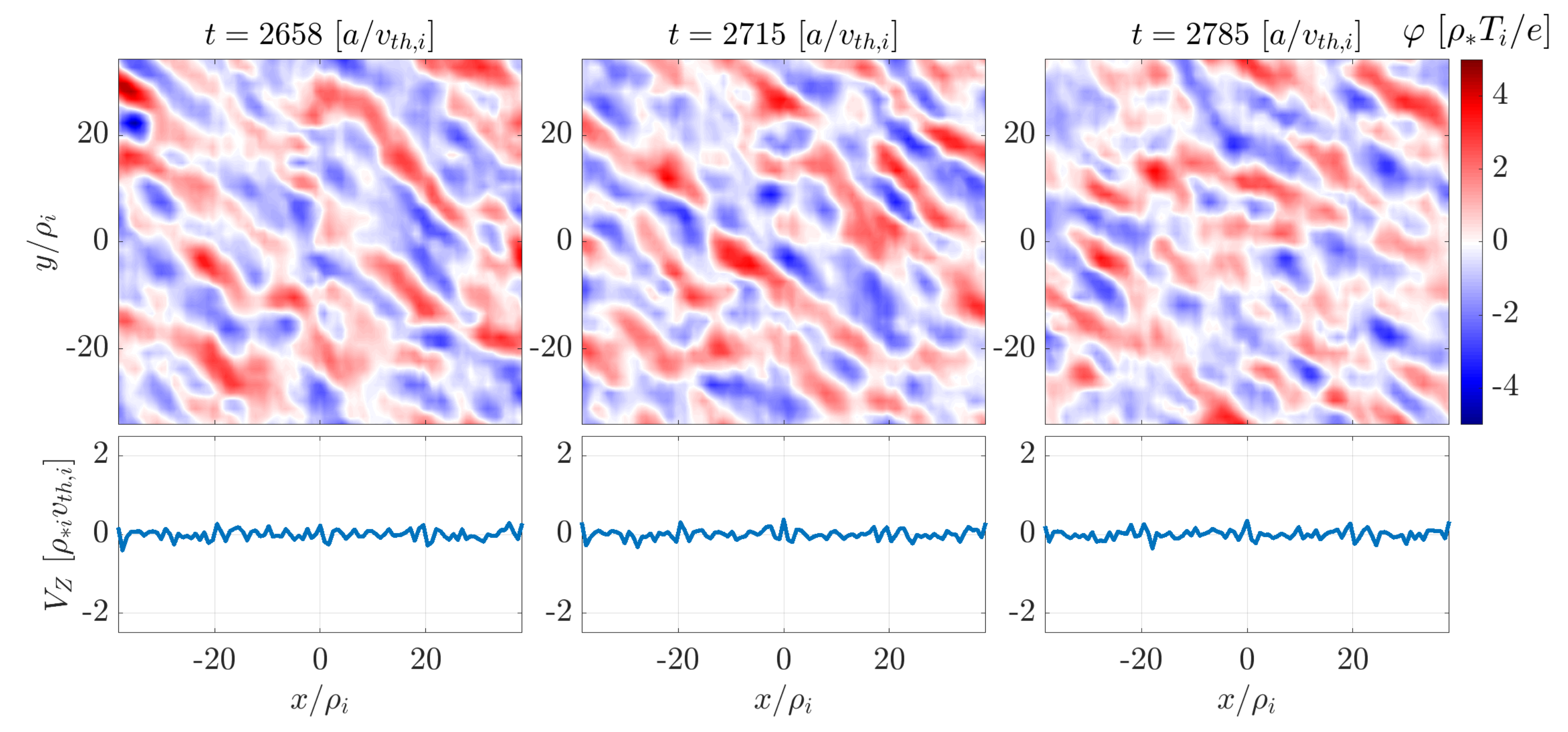}
\caption{Consecutive snapshots of the turbulence in real space for the low-transport state where $\gslength/\LTi = 1.76$ and $\gammaE=-0.079$. In the top panels, the fluctuating elecrostatic potential is plotted at three successive times at the outboard of the torus, in the plane perpendicular to $\bm{B}$. In the bottom panels, the zonal flow is plotted at the same times.}
\label{fig:real_space_low_state}
\end{figure}

\begin{figure}
\centering
\includegraphics[scale=0.38]{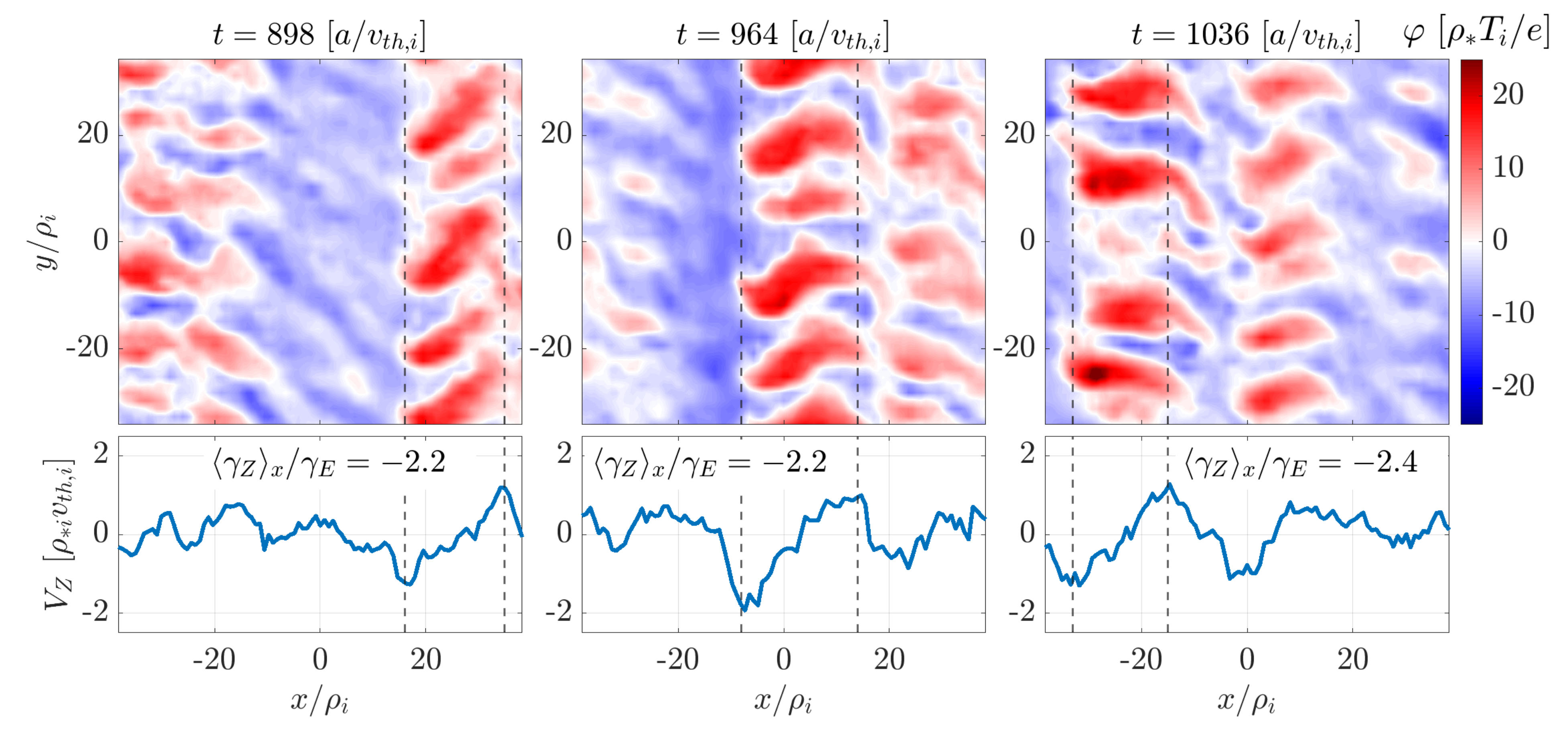}
\caption{Same as figure \ref{fig:real_space_low_state} but for the high-transport state. In the bottom panels, the zonal shear averaged between the two vertical dashed lines is compared to the externally imposed mean flow shear.}
\label{fig:real_space_high_state}
\end{figure}

The two states observed here are distinguished by significant differences in the amplitudes of their turbulent fluctuations and by the spatial structure of the turbulence. In figure \ref{fig:real_space_low_state}, we show a typical snapshot of turbulence in a low-transport state. The contours of the fluctuating electrostatic potential are plotted in the plane perpendicular to the magnetic field at the outboard of the torus. The $x$ coordinate measures the distance along the normal to a magnetic-flux surface and the $y$ coordinate labels the magnetic field lines within the surface. The simulation is done in the frame moving with the mean flow at $x=0$: the $y$-component of the mean flow thus has the opposite sign as $x$. The turbulent eddies feature a clear tilt as they are being sheared by the mean flow, similarly to \cite{shaferPoP12, vanwykJPP16, vanwykPPCF17, foxPPCF17}. In figure \ref{fig:real_space_high_state}, we show consecutive snapshots of turbulence with the same equilibrium parameters as in figure \ref{fig:real_space_low_state}, but in the high-transport state: bands of high-amplitude eddies propagate radially across the simulation domain, and eddies do not feature any clear tilt. This intermittent high-transport state is reminiscent of advecting structures reported in \cite{mcmillanPoP09, mcmillanJPP18, ajayThesis20}.

\section{The role of zonal modes}

We find that the presence of ``zonal'' modes is a crucial distinguishing feature between the two states. Modes are called zonal when they have no spatial variation other than in the radial ($x$) direction. They are linearly stable and cannot be sheared by a toroidal mean flow, but they can exchange energy with non-zonal modes via nonlinear interactions. Zonal modes include zonal flows with a shearing rate $\gammaz$, which can affect the rest of the turbulence in a manner analogous to the mean flow shear $\gammaE$. The zonal flows are known to develop through a secondary instability of the modes driven unstable by the temperature gradient \cite{rogersPRL00, diamondPPCF05, ivanovJPP20}.

When the amplitudes of zonal modes become large enough, we find that the zonal shear can compete with, and indeed obviate, the mean flow shear. Such a negation of the equilibrium shear by a zonal shear was already explored in previous work, e.g., \cite{mcmillanPoP09, mcmillanJPP18}. In the lower panels of figure \ref{fig:real_space_high_state}, we plot the zonal flow $\vzonal$ ($\propto \partial\phiturbz / \partial x$ where $\phiturbz$ is the zonal part of the potential). We observe radially-propagating bands within which the zonal shear $\gammaz$ ($\propto \partial\vzonal / \partial x$) is of the same order of magnitude as the background shear $\gammaE$, but carries the opposite sign. In such bands where the zonal and mean shears oppose each other, non-zonal fluctuations grow faster and feed the zonal modes nonlinearly, until the system settles in the high-transport state. We also show in figure \ref{fig:hysteresis} that the transport obtained in the complete absence of mean flow shear (green `+' symbols) is much closer to the high-transport states than to the low-transport states.

\begin{figure}
\centering
\includegraphics[scale=0.55]{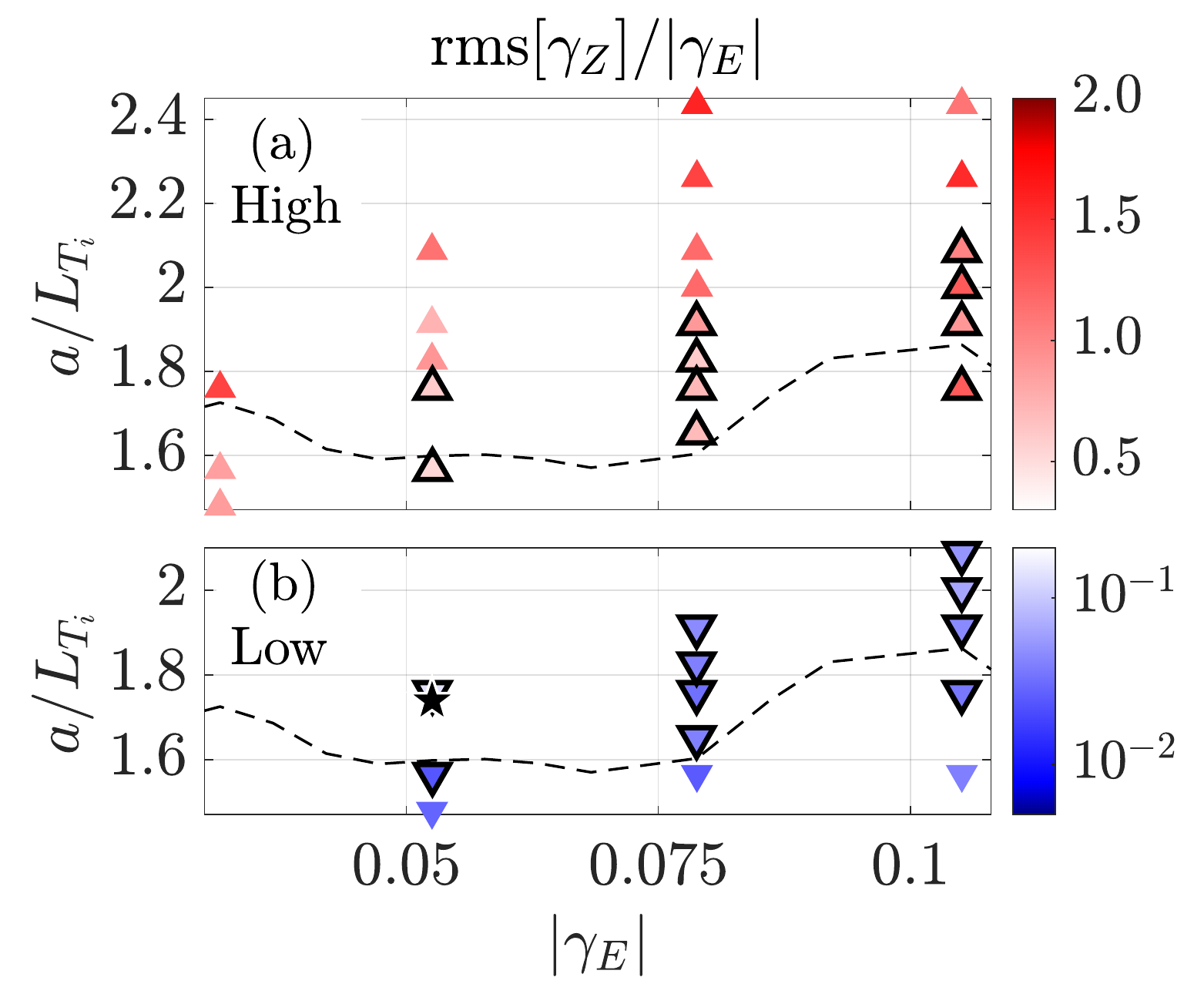}
\caption{RMS zonal shear versus $\gammaE$ for the two turbulent states. High-transport states are shown in panel (a), and low-transport states in panel (b). Black-bordered markers indicate parameters at which either a high-transport or a low-transport state can be obtained, depending on the initial size of fluctuation amplitudes. The parameters of the experiment considered here are shown by a black star in panel (b). The dashed line marks the temperature gradient as a function of $\gammaE$ below which turbulence is subcritical ($\gammaavg < 0$).
}
\label{fig:gammaz_v_gexb}
\end{figure}

In low-transport states, zonal modes do not seem to play a crucial role for the turbulent dynamics: unlike in the high-transport states, no long-lived structures with $\gammaz$ opposing $\gammaE$ are observed. Further evidence of the weaker impact of zonal modes in low-transport states can be seen in simulations where we artificially set zonal modes to zero at every time step, indicated by black crosses in figure \ref{fig:hysteresis}. Despite this unphysical truncation introduced in the system, and independently of the initial condition, a saturated state is obtained that is -- apart from a slight change in the flux -- indistinguishable from the low-transport state.

The occurrences of low-transport and high-transport states for a range of mean flow shear rates are shown in figure \ref{fig:gammaz_v_gexb}, where we plot the ratio $\textrm{rms}[\gammaz]/\gammaE$. Here, $\textrm{rms}[\gammaz] = \sqrt{\langle\phiturbz^2\rangle_{t,x}}/\lcorrxzonal^2$ and we denote by $\lcorrxzonal$ the radial correlation length of the zonal modes, which we define in appendix \ref{sec:corr_time}. As a result of the interplay between the zonal modes and the mean flow, high-transport states are only obtained when the initial fluctuation amplitudes are sufficiently large, or when the fluctuation amplitudes become large enough for the zonal shear to start competing with $\gammaE$ (e.g., when $\gslength/\LTi$ is increased or $\gammaE$ is decreased past a certain threshold). Panel (a) of figure \ref{fig:gammaz_v_gexb} confirms that the magnitude of the zonal shear in the high-transport states is comparable to that of the mean shear. Panel (b) indicates that low-transport states only survive when the zonal shear is much smaller than $\gammaE$ (roughly by an order of magnitude). This last result could be due to a feedback mechanism whereby even a weak $\gammaz$ can partially oppose $\gammaE$, allowing for larger turbulent amplitudes -- and therefore larger zonal modes -- to develop.

\section{Two correlation time scales}

\begin{figure}
\centering
\includegraphics[scale=0.55]{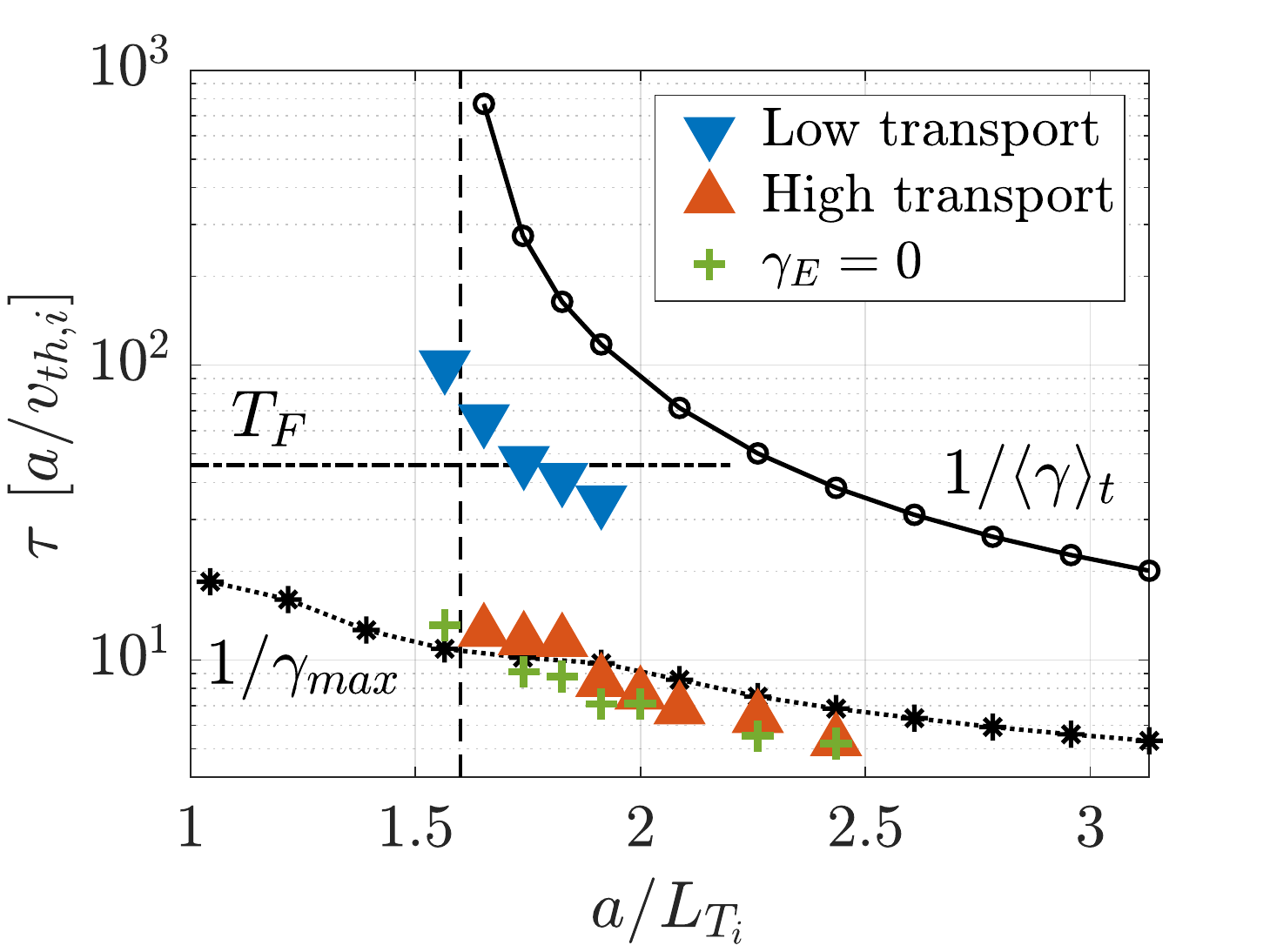}
\caption{Correlation time of turbulent eddies in the low-transport state, high-transport state, and in the absence of a mean flow shear. The vertical dashed line marks the temperature gradient below which there is no effective instability in the presence of a mean flow shear, i.e., $\gammaavg \leq 0$. In the simulations labelled by green `+' signs, the externally imposed mean flow shear was set to zero. For all other simulations, $\gammaE=-0.079$ was used. Considering other $\gammaE$ values leads to similar results.}
\label{fig:tcor}
\end{figure}

In a saturated turbulent state, the correlation time of the turbulence can be estimated from the gyrokinetic equation by $\taunl = \left[c \rms{\phiturbnz} / (B \lcorrx \lcorry)\right]^{-1}$, where $\rms{\phiturbnz}$ is the RMS value of the non-zonal electrostatic potential, $\lcorry$ is the eddy correlation length in the $y$ direction (defined in appendix \ref{sec:corr_time}), $c$ is the speed of light and $B$ is the magnetic field strength. For high-transport states, figure \ref{fig:tcor} shows that $\taunl \sim 1/\gammamax$, where $\gammamax$ is the maximum instantaneous linear growth rate in the presence of flow shear (close to the linear growth rate in the absence of flow shear). For low-transport states, $1/\gammamax < \taunl < 1/\gammaavg$, which may suggest that the average linear growth rate plays a role in setting the saturated turbulent amplitudes in those states. In order for $\gammaavg$ to be relevant in a turbulent state, eddies must be able to survive longer than a Floquet period, i.e. $\taunl \gtrsim \Tfloq$, as is approximately the case for the low-transport states in figure \ref{fig:tcor}. While the competition between zonal and mean flow shear is likely a generic feature of magnetised plasma turbulence, the existence of $\gammaavg \neq \gammamax$ requires toroidicity. Further studies are needed to determine if these two distinct growth rates are a necessary feature of the bistable turbulence reported here -- and thus if similar bi-stable states are likely to be found beyond toroidal plasmas.

\section{Consequences of bistability} \label{sec:consequences}

The bistability reported in this work may lead to the existence of bifurcations. As we have argued, low-transport states cease to exist when the fluctuation amplitudes increase past a certain threshold value. If we now consider a plasma in which, instead of being fixed, the temperature gradient is slowly increasing in time, a discontinuous jump will be triggered from a low-transport state to the high-transport branch. The same jump can be triggered by decreasing the mean flow shear in a low-transport state. As shown in figure \ref{fig:gammaz_v_gexb}, we observe that the subcritical low-transport states exist closer to marginal stability in the $(\gammaE,\gslength/\LTi)$ plane than the subcritical high-transport states. We attribute this to the intermittent nature of the high-transport state associated with the radially propagating bands shown in figure \ref{fig:real_space_high_state}. Previous work with neutral fluid flows \cite{faisstJFM04}, accretion disks \cite{rempelPRL10} and fusion plasmas \cite{highcockPoP11, barnesPRL11, vanwykPPCF17} has indeed shown that the survival of subcritical turbulence over long times is compromised by rare, large fluctuations. Similarly to the transition from low to high transport, we argue that subcritical high-transport states can drop to the low-transport branch if the temperature gradient slowly decreases in time. From figure \ref{fig:gammaz_v_gexb}, we expect that the same transition could be achieved by increasing the mean flow shear in a subcritical high-transport state.

\begin{figure}
\centering
\includegraphics[scale=0.55]{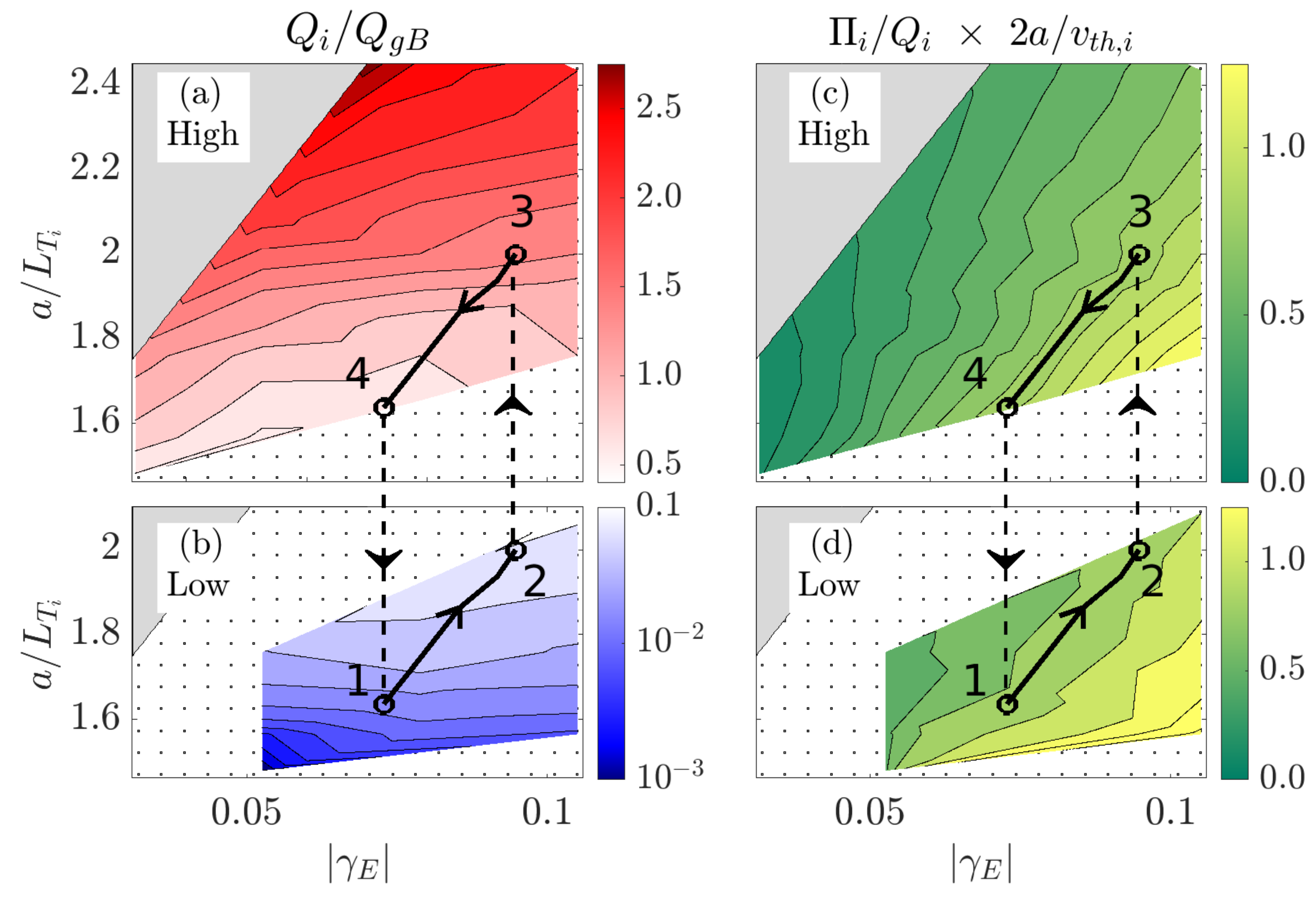}
\caption{Dependence of the ion heat flux (a,b) and the momentum-to-heat-flux ratio (c,d) on the imposed flow shear and the inverse ion-temperature-gradient scale length. The top panels show results for the high-transport states, the bottom panels for the low-transport states. Dotted areas in the upper (respectively lower) panels indicate areas where no high-transport (respectively low-transport) state could be obtained. The grey areas indicate parameter ranges where no simulations were run. There is a gap between the values of the heat flux obtained in (a) and those obtained in (b). The path defined by points A, B, C and D gives an example of the successive stages of a gradient-relaxation cycle, when the heat injected into the plasma corresponds to a flux within the aforementioned gap.}
\label{fig:fluxes_contour}
\end{figure}

Existence of bifurcations opens up the possibility for relaxation cycles of the mean gradients to develop. This hinges on two findings that we show in figure \ref{fig:fluxes_contour}. First, we observe a significant gap between the highest heat flux obtained in low-transport states and the lowest heat flux obtained in high-transport states. Second, we observe that the ratio of the turbulent momentum flux to heat flux, $\Pi_i/Q_i$, is almost identical in the low-transport and high-transport states.

We now consider a thought experiment in which an external power $P$ is injected into the volume bounded by a given magnetic-flux surface of area $S$, via a beam of neutral atoms with energy $E$. As is argued in \cite{parraPRL11}, the turbulent heat flux exiting the magnetic-flux surface is $Q_i \sim P/S$, and $\Pi_i/Q_i \sim E^{-1/2} \vthi/\gslength$. Thus, a given beam configuration corresponds to a unique pair $(Q_i, \Pi_i/Q_i)$. We consider an initial situation where the input power is such that $(P,E)$ corresponds to the levels of turbulent fluxes of a low-transport state (point A in figure \ref{fig:fluxes_contour}). From this initial stationary state, we increase $P$ by small successive increments, keeping $E$ fixed. In response, the plasma equilibrium will evolve through a succession of low-transport stationary states with ever larger $Q_i$, but with $\Pi_i/Q_i$ staying constant (along the solid arrow up to point B in figure \ref{fig:fluxes_contour}).

Above a certain threshold, we find that there is a range of values of $P$ with no corresponding solutions for the turbulent fluxes: in figure \ref{fig:fluxes_contour}, these are the powers too high to match the low-transport state at point B, but too low to match the high-transport state at point C. It is then interesting to ask what will happen to a plasma where the input power falls into this gap. One (unexciting) possibility is that an actual solution may exist outside of the region of parameter space explored here, and that the plasma migrates to that solution. Another (more interesting) possibility would be for the temperature gradient to continue increasing until the plasma transitions to a high-transport state (jumping from B to C in figure \ref{fig:fluxes_contour}). In this state, the outgoing heat flux is larger than what can be sustained by the external power input, and the temperature gradient starts to flatten. As $\gslength/\LTi$ decreases (from C to D), the turbulent fluctuations remain too large to allow a transition back to a low-transport state. Eventually, $\gslength/\LTi$ becomes too small for the high-transport state to survive, and the system transitions back to the lower state (from D to A). The flux is now too low compared to the power input, so the temperature gradient builds up again, and the cycle repeats itself. In this scenario, no proper steady state is reached when the input power falls within a ``forbidden'' gap, and the temperature gradient and mean flow shear would experience periodic relaxation cycles.


\section{Discussion} \label{sec:discussion}

We have found that near-marginal turbulence in fusion devices is bistable, and regulated by the competition between external shear and zonal modes. The existence of bistability suggests a new approach to long-standing questions around bifurcations and gradient relaxation cycles observed in fusion devices \cite{vongoelerPRL74, wagnerPRL82, hastieASS97, connorPPCF98}.
%
This work also presents a new challenge for a research area where the prevailing assumption has been a one-to-one correspondence between plasma parameters and turbulent transport.
%
Further work could focus on how the extent of the bistable region might be modified, for example by exploring the effect of collisions on the saturation of zonal modes \cite{colyerPPCF17, weiklPoP17}. Another avenue of interest may be to determine how bistability manifests itself in flux-driven gyrokinetic simulations. Experiments could test the existence of bistability in fusion devices, following a scenario similar to the one described in figure \ref{fig:fluxes_contour}. The idea of relaxation cycles discussed in section \ref{sec:consequences} could be considered in the context of subcritical turbulence, where transitions might occur from a situation with no turbulent transport to a state with a finite level of turbulent transport. Lastly, we note that the details of the plasma analysed here, such as the exact nature of the drive for turbulence and perhaps toroidicity, do not appear to be crucial to our understanding of bistable states: the only requirements we have identified so far are an applied flow shear and the ability of the plasma to generate zonal flows. Therefore, we expect that similar effects may be observed in a variety of systems.


The authors are grateful to H. Weisen and P. Sir\'en for providing the experimental data used in this work. They especially thank F. Parra for his insightful comments on the manuscript. They would also like to thank O. Beeke, J. Parisi and J. Ruiz Ruiz for very fruitful discussions. NC was supported by a Berrow Foundation Scholarship, the Steppes Fund for Change and the Fondation H\'el\`ene et Victor Barbour. The work of MH was funded by EPSRC grant EP/R034737/1, as was, in part, the work of MB and AAS. Computing resources were provided on the ARCHER High Performance Computer through the Plasma HEC Consortium, EPSRC grant EP/L000237/1 under project e607, on the EUROfusion High Performance Computer (Marconi-Fusion) under the projects FUA34\_MULTEI and FUA35\_OXGK, and on the JFRS-1 supercomputer system at the International Fusion Energy Research Centre's Computational Simulation Centre (IFERC-CSC) at the Rokkasho Fusion Institute of QST (Aomori, Japan) under the project MULTEI.

The authors report no conflict of interest.


\begin{appendices}

\section{Gyrokinetic system} \label{sec:gk_system}

In this work, we follow the $\delta f$ gyrokinetic approach \cite{cattoPP78, friemanPoF82, sugamaPoP98, abelRPP13}, which relies on the scale separations present in the plasma to describe the time evolution of turbulent fluctuations. The ratio of gyroradius to machine size $\rhostarspec = \gradius_\spec/\gslength \ll 1$ for species $\spec$ is used as the asymptotic-expansion parameter. The minor radius of the device is denoted by $\gslength$ and the gyroradius is given by $\gradius_\spec = \vert \bunit\times\bm{v}/\Omega_\spec\vert$, where $\bunit$ is the unit vector in the direction of the magnetic field $\bm{B}$ and $\bm{v}$ is the velocity of the particle. The gyrofrequency of the particle is $\Omega_\spec=\charge_\spec B/m_\spec c$, where $\Znum_\spec$ and $m_\spec$ are, respectively, the charge number and mass of the particle, $e$ is the elementary charge, and $c$ is the speed of light. The amplitudes of the fluctuations are ordered to be $\bigO{\rhostarspec}$ smaller than the corresponding mean quantities. The turbulent time scale is ordered to be $\bigO{\rhostarspec^2}$ shorter than the time scale of the evolution of mean plasma parameters, but $\bigO{\rhostarspec^{-1}}$ longer than the Larmor periods of the particles. Moreover, it is assumed that fluctuations can stretch far along magnetic field lines, but that they only span a few gyroradii across field lines. The orderings in time and space can be summarised as
\be \label{eq:orderings_time}
\frac{d}{dt}\ln(\deltaf_\spec) \sim \rhostarspec^{-2}\frac{d}{dt}\ln(F_\spec) \sim \bigO{\rhostarspec\Omega_\spec},
\ee
\be \label{eq:orderings_space}
\bunit\cdot\nabla\ln(\deltaf_\spec) \sim \rhostarspec\nrm{\nabla\ln(\deltaf_\spec)} \sim \nrm{\nabla\ln(F_\spec)} \sim \bigO{1/\gslength},
\ee
where $\deltaf_\spec$ is the fluctuating part of the distribution function of particles and $F_\spec$ is their mean distribution function (averaged over the turbulent time scales and over the turbulent length scales across $\bm{B}$). Here, $d/dt = \partial/\partial t + \bm{\uflow}\cdot\nabla$ is the convective time derivative with respect to the mean flow $\bm{\uflow}$.

The geometry of the system considered here is typical of magnetic-confinement-fusion experiments. The plasma is confined in a toroidally shaped magnetic cage. The magnetic field lines of this cage wind around the torus, tracing out nested toroidal surfaces, known as magnetic-flux surfaces. The rapid gyromotion about magnetic field lines limits the ability of charged particles to move across these flux surfaces.

We focus on plasmas with a mean flow such that $\rhostarspec \ll \nrm{\bm{\uflow}} / \vthi \ll 1$, where $\vthi = \sqrt{2T_i/m_i}$ is the ion thermal speed and $T_i$ is the ion temperature multiplied by the Boltzmann constant $\boltzcst$. In this ``intermediate-flow'' ordering, we can neglect the centrifugal force, and the mean flow is purely toroidal \cite{cattoPoF87, abelRPP13}: $\bm{\uflow}=\torfreq R^2 \nabla\torang$ with $\torfreq$ the angular rotation frequency, $R$ the major radius of the torus and $\torang$ the toroidal angle. It follows that the perpendicular and parallel flow shear rates are related by a geometrical factor, and both can be expressed in terms of the shearing rate $\gammaE = (\rpsimid/q_\midline) \partial\torfreq/\partial\rpsi \rvert_{\rpsimid}$, where the subscript `$\midline$' denotes quantities evaluated on the flux surface of interest, $\rpsi$ is the half-width of the flux surface at the height of the magnetic axis and the safety factor $q=(2\pi)^{-1}\int_0^{2\pi} d\polang (\bm{B}\cdot\nabla\torang)/(\bm{B}\cdot\nabla\polang)\vert_\magflux$ is the number of toroidal turns required by a field line to wind once around the torus poloidally. The magnetic shear appearing in section \ref{subsec:model} is defined as $\shat = (\rpsimid/q_\midline) \partial q/\partial\rpsi \vert_{\rpsimid}$. We further restrict consideration to cases with low thermal-to-magnetic-pressure ratio (plasma beta) and hence only retain electrostatic fluctuations. We neglect all effects associated with impurities in the plasma, and only consider electrons and the main hydrogenic ion species.

After averaging over the rapid gyromotion of particles, the gyrokinetic equation can be written as
\begin{multline} \label{eq:gk}
\frac{d\gavg{\deltaf_\spec}}{dt}
+\left( \relvel_\parallel\bunit+\vbs+\vcs+\gavg{\veturb} \right) \cdot\nabla \left( \gavg{\deltaf_\spec} + \frac{\charge_\spec\gavg{\phiturb}}{T_\spec}\eqf \right) = \\
\gavg{C[\deltaf_\spec]}- \gavg{\veturb} \cdot\left( \frac{\rmaj B_\torang}{B} \frac{m_\spec \relvel_\parallel}{T_\spec} \eqf \nabla \torfreq + \nabla \eqf \right)
\end{multline}
in $(\gcenter, \energy, \magmom, \gang)$ coordinates, where $\gcenter = \ppos - \bm{\gradius}_\spec$ is the particle's gyrocenter, $\ppos$ is its position, $\energy=m_\spec \relvel^2/2$ is its kinetic energy, $\magmom=m_\spec \relvel_\perp^2/2B$ its magnetic moment and $\gang$ its gyrophase. 
Here, $\gavg{\cdot}$ denotes an average over $\gang$ at fixed $\gcenter$, $\phiturb$ is the fluctuating electrostatic potential, $\bm{\relvel}$ is the particle velocity relative to $\bm{\uflow}$, subscripts $\parallel$ and $\perp$ indicate components along and across $\bm{B}$ respectively, $\eqf$ is a local Maxwellian velocity distribution, $C$ is the collision operator and $B_\torang$ is the toroidal component of $\bm{B}$. The drift velocity due to magnetic curvature and $\nabla B$ is $\vbs = \bunit/\Omega_\spec\times [ \relvel_\perp^2\nabla\ln(B)/2 + \relvel_\parallel^2 \bunit\cdot\nabla\bunit ]$, and the Coriolis drift velocity is $\vcs = (2\relvel_\parallel \torfreq/\Omega_\spec)\bunit\times(\hat{\bm{z}}\times\bunit)$ with $\hat{\bm{z}}$ the unit vector in the vertical direction. The nonlinearity in equation \eqref{eq:gk} stems from the fluctuating $\bm{E}\times\bm{B}$ drift $\veturb = c\bunit/B\times\nabla\phiturb$ advecting $\deltaf_\spec$ on the left-hand side.
Perpendicular flow shear enters via the convective time derivative, while the drives from the shear in the parallel flow and the temperature gradient, respectively, enter via the $\nabla\torfreq$ and $\nabla\eqf$ terms on the right-hand side. The temperature gradient is specified by the normalised inverse gradient length $\gslength/\LTs=-\gslength\ d(\ln T_\spec)/d\rpsi$.
The set of equations is closed by the quasineutrality condition:
\be \label{eq:qn}
\sum_\spec \Znum_\spec \int\dd^3 \relvel \ringavg{\gavg{\deltaf_\spec}} = \sum_\spec \frac{e\Znum_\spec^2}{T_\spec}\left( n_\spec\phiturb - \int\dd^3 \relvel 
\ringavg{\gavg{\phiturb}} \eqf \right),
\ee
where $n_\spec$ is the particle density and $\ringavg{\cdot}$ denotes an average over $\gang$ at fixed particle position.

Fluctuations with no spatial variation other than in the radial direction are known as ``zonal'' fluctuations, and produce sheared $\bm{E}\times \bm{B}$ drifts in the $y$ direction. The zonal flow is given by
\be \label{eq:zonal_flow}
\vzonal = -\frac{c}{B}\lvert \bunit\times\nabla\rpsi\rvert\frac{\partial\phiturbz}{\partial\rpsi},
\ee
where $\phiturbz$ is the zonal part of the electrostatic potential. The shear of the zonal flow is $\gammaz = \partial\vzonal / \partial\rpsi$.

The system of equations \eqref{eq:gk} and \eqref{eq:qn} is solved for $\deltaf_\spec$ and $\phiturb$ using the local gyrokinetic code \tw{GS2} \cite{kotschenreutherCPC95, barnesPoP09, highcockThesis12, christenJPP21} in a filament-like simulation domain (known as a flux tube \cite{beerPoP95}) that follows a magnetic field line around the flux surface of interest. The flux-surface label $x = (q_\midline/\rpsimid B_\rfr)(\magflux-\magflux_\midline)$ and the field-line label $y = (1/B_\rfr)(\partial\magflux/\partial\rpsi)\rvert_{\rpsimid}(\alpha-\alpha_\midline)$ are used as coordinates across $\bm{B}$. The poloidal angle $\polang$ serves as the coordinate along $\bm{B}$. Here, $B_\rfr$ is a reference magnetic field strength, $\magflux = \int_{0}^{r} dr' r'R\bm{B}\cdot\nabla\polang$ is the poloidal magnetic flux, $r$ is the minor radius of the torus, and $\alpha=\torang-\int_0^\polang d\polang^\prime (\bm{B}\cdot\nabla\torang)/(\bm{B}\cdot\nabla\polang)\vert_\magflux$ labels field lines on a given flux surface. The code computes the turbulent contribution to the heat and momentum fluxes given by
\be \label{eq:heatflux}
Q_\spec = \flxsurfavgbig{\int\dd^3\bm{\relvel} \frac{m_\spec \relvel^2}{2} \deltaf_\spec \veturb \cdot \nabla\magflux},
\ee
\be \label{eq:momflux}
\Pi_\spec = \flxsurfavgbig{m_\spec \rmaj^2 \int\dd^3\bm{v} \left(\bm{v}\cdot\nabla\torang\right)\deltaf_\spec \veturb \cdot \nabla\magflux},
\ee
respectively, with $\flxsurfavg{\cdot}$ denoting the volume average over the flux tube.

\section{Correlation time and correlation lengths} \label{sec:corr_time}

\begin{figure}
\centering
\includegraphics[width=\textwidth]{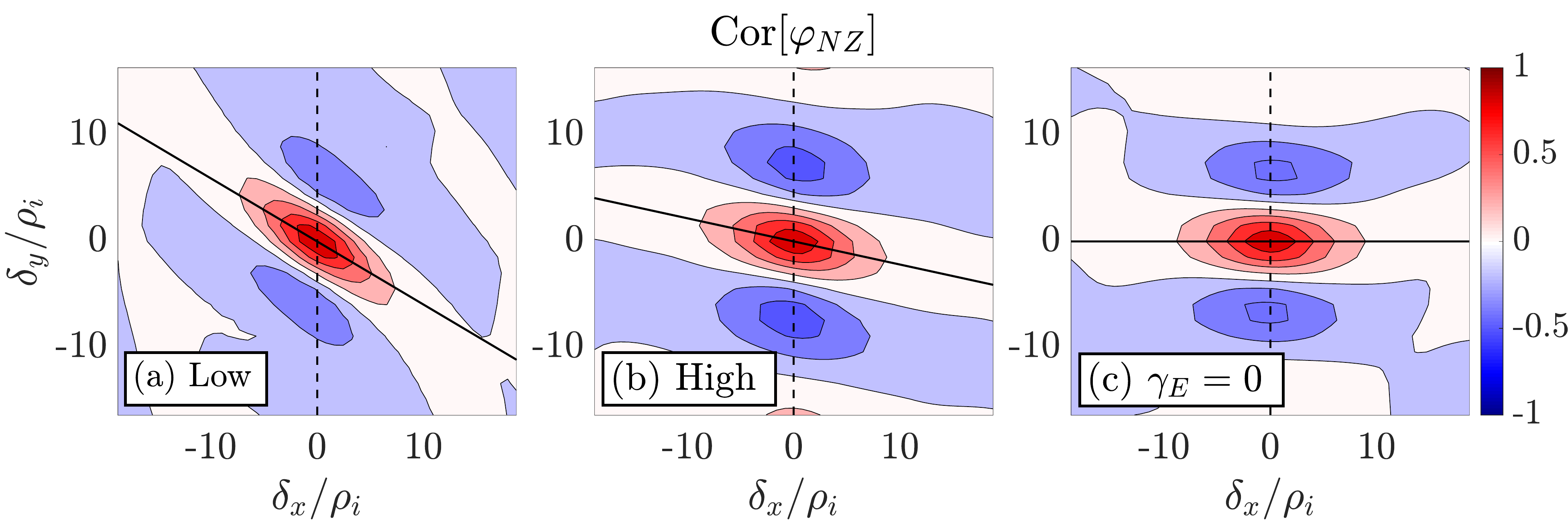}
\caption{Two-point spatial correlation function for a low-transport state (a), a high-transport state (b) and a state with no mean flow shear (c). For the three states, $\gslength/\LTi=1.76$. In (a) and (b), $\gammaE = -0.079$.}
\label{fig:corrxy}
\end{figure}

Given a saturated turbulent state, we estimate the eddy correlation time as being
\be \label{eq:tcorr_def}
\taunl = \left[\frac{c}{B}\frac{\rms{\phiturbnz}}{\lcorrx\lcorry}\right]^{-1},
\ee
where $\lcorrx$ and $\lcorry$ denote the eddy correlation length in the $x$ and $y$ direction, respectively, and where $\rms{\phiturbnz} = \sqrt{\langle\phiturbnz^2\rangle_{t,x,y}}$ is the root mean square of the nonzonal part of the electrostatic potential $\phiturbnz$, averaged over time, $x$ and $y$. The expression \eqref{eq:tcorr_def} is obtained from the nonlinear term in the gyrokinetic equation \eqref{eq:gk}. We then define the two-point spatial correlation function
\be \label{eq:corrfunc_def}
\corrfunc[\phiturb](\deltax,\deltay) = \frac{\big\langle\phiturb(t,x,y)\phiturb(t,x+\deltax,y+\deltay)\big\rangle_{t,x,y}}{\big\langle\phiturb^2(t,x,y)\big\rangle_{t,x,y}^{1/2}\big\langle\phiturb^2(t,x+\deltax,y+\deltay)\big\rangle_{t,x,y}^{1/2}}.
\ee
The correlation lengths $\lcorrx$ and $\lcorry$ are chosen to correspond to the $e$-folding of $\corrfunc[\phiturbnz]$ along the $\deltax$ direction (adjusted to match the tilt induced by the flow shear) and the $\deltay$ direction, respectively. Typical examples of $\corrfunc[\phiturbnz]$ are shown in figure \ref{fig:corrxy}. The zonal correlation length $\lcorrxzonal$ corresponds to the $e$-folding of $\corrfunc[\phiturbz]$ along the $\deltax$ direction. Note that the exact definition of the correlation lengths is somewhat arbitrary. Another choice, that is commonly found in the literature, is to define $\lcorrx$ and $\lcorry$ as the integral of $\corrfunc[\phiturbnz]$ along the $\deltax$ and $\deltay$ axes -- which, in our case, yields similar results to the $e$-folding lengths.

\section{Numerical parameters}

The authors would like to thank Henri Weisen and Paula Sir\'en for providing the experimental data discussed here. In this letter, we focus on the plasma discharge \#68448 carried out at the JET tokamak. This discharge is of interest for two reasons: it features a sheared mean toroidal flow and it is well diagnosed. The discharge is documented in the \tw{JETPEAK} database \cite{sirenJI19}. Plasma parameters at $\rpsi/\gslength = 0.51$ are presented in table \ref{tab:jet_param}. Only the main ion species (deuterium) is considered. Electromagnetic effects are neglected.

\begin{table}
\centering
\begin{tabular}{c|c|c}
$I_{\text{p}}$ & $2.6$MA & Plasma current \\
$B_{\text{T}}$ & $2.9$T & Vacuum toroidal field on axis \\
$P_{\text{NBI}}$ & $17$MW & Neutral beam heating power \\
$\Rmaj$ & $3.06\gslength$ & $\left[\max(R)+\min(R)\right]/2$ for this flux surface \\
$\rpsi$ & $0.508\gslength$ & $\left[\max(R)-\min(R)\right]/2$ for this flux surface \\
$\lvert q_\midline\rvert$ & $1.43$ & flux-surface averaged safety factor \\
$\shat$ & $0.574$ & flux-surface averaged magnetic shear \\
$\elong$ & $1.36$ & Miller elongation \cite{millerPoP98} \\
$\dd\elong/\dd\rpsi$ & $0.146/\gslength$ & elongation gradient \\
$\triang$ & $0.0571$ & $\arcsin$ of Miller triangularity \cite{millerPoP98} \\
$\dd\triang/\dd\rpsi$ & $0.129/\gslength$ & gradient of \tw{GS2} triangularity \\
$\gammaE$ & $-0.0553v_{\text{th},i}/\gslength$ & background flow shear rate \\ 
$\torfreq$ & $-0.08v_{\text{th},i}/\gslength$ & background flow angular frequency \\
$n_i/n_e$ & $1.0$ & ion to electron density ratio \\
$1/\Lni$ & $0.602/\gslength$ & inverse ion density gradient length \\
$1/\Lne$ & $0.602/\gslength$ & inverse electron density gradient length \\
$T_e/T_i$ & $0.855$ & electron to ion temperature ratio \\
$1/\LTi$ & $1.7392/\gslength$ & inverse ion temperature gradient length \\
$1/\LTe$ & $1.551/\gslength$ & inverse electron temperature gradient length \\
$\nu_{ii}$ & $2.6\times10^{-4}v_{\text{th},i}/\gslength$ & ion collisionality \\
$\nu_{ee}$ & $0.02v_{\text{th},i}/\gslength$ & electron collisionality \\
$\beta$ & $0.0125$ & $2 e \mu_0 n_i T_i / B_r^2$ \\
\end{tabular}
\caption{Parameters for the JET discharge \#68448 at $\rpsi/\gslength=0.51$. The gradient length of a given quantity $\xi$ is defined as $L_\xi=1/\left[\dd\log(\xi)/\dd\rpsi\right]$. $\mu_0$ denotes the vacuum permeability.}
\label{tab:jet_param}
\end{table}

In order to ensure that the results presented in this letter are not affected by insufficient numerical resolution, scans were carried out for various numerical parameters of the gyrokinetic code \tw{GS2} \cite{kotschenreutherCPC95, barnesPoP09, highcockThesis12, christenJPP21}. The scanned values are presented in table \ref{tab:scan_params}.

Below, we also provide a typical \tw{GS2} input file used for this letter, based on the experimental data of the JET discharge \#68448. To obtain a low-transport state, the simulation should be started with the mean flow shear turned on, i.e., with \tw{g\_exb} set to the value given below. To obtain a high-transport state, the simulation can be started with no flow shear (\tw{g\_exb} set to zero) until a saturated state is reached, and it should then be restarted (instructions are given in the comments below) with the flow shear turned on.

\begin{table}
\centering
\begin{tabular}{c|c|c|c|c}
Parameter 	& Values tested & Type of scan 	& Value used 	& Units \\
\hline
$\Deltakx$ 	& 0.04 -- 0.08 	& nonlinear 	& 0.08			& $1/\gradius_i$ \\
$\Kx$ 		& 1.7 -- 15.2 	& nonlinear 	& 3.8			& $1/\gradius_i$ \\
$\Deltaky$ 	& 0.045 -- 0.18 & nonlinear 	& 0.09			& $1/\gradius_i$ \\
$\Ky$ 		& 0.99 -- 1.98 	& nonlinear 	& 1.98			& $1/\gradius_i$ \\
\tw{ntheta} & 16 -- 128 	& linear 		& 32			& - \\
\tw{negrid} & 6 -- 48 		& linear 		& 16			& - \\
\tw{ngauss} & 3 -- 20 		& linear 		& 5				& - \\
\tw{vcut} 	& 2.5 -- 4.5 	& linear 		& 2.5			& - \\
$\Delta t$ 	& 0.025 -- 0.1 	& linear 		& linearly: 0.1	& $\gslength / \vthi$ \\
\end{tabular}
\caption{Ranges of numerical parameters that were tested. Here, \tw{ntheta} roughly denotes the number of grid points in $\polang$, \tw{negrid} denotes the number of energy grid points, $4\times$\tw{ngauss} the number of untrapped pitch angles, and \tw{vcut} the number of standard deviations from the Maxwellian distribution of velocities above which the fluctuating distribution function is set to zero}
\label{tab:scan_params}
\end{table}

\newpage

\begin{lstlisting}
!! ------------------------------------------------- !!
!!  GS2 input file based on the JET discharge #68448  !!
!! ------------------------------------------------- !!


&species_knobs
 nspec = 2
/

&species_parameters_1
 z = 1.0
 mass = 1.0
 dens = 1
 temp = 1
 tprim = 1.91312 ! corresponds to a/LTi
 fprim = 0.60228
 uprim = 0.0
 vnewk = 0.00026042
 type = 'ion'
/

&species_parameters_2
 z = -1.0
 mass = 2.7e-4
 dens = 1
 temp = 0.85478
 tprim = 1.5509
 fprim = 0.60228
 uprim = 0.0
 vnewk = 0.019972
 type = 'electron'
/

&dist_fn_species_knobs_1
 fexpr = 0.45
 bakdif = 0.05
/

&dist_fn_species_knobs_2
 fexpr = 0.45
 bakdif = 0.05
/

&collisions_knobs
 collision_model='default'
/

&parameters
 beta = 0.0
 zeff = 1
/

&theta_grid_parameters
 ntheta = 32
 nperiod = 1
 rhoc = 0.50825
 shat = 0.57383
 qinp = -1.4253
 Rmaj = 3.0642
 R_geo = -3.0642
 shift = -0.10502
 akappa = 1.3594
 akappri = 0.1458
 tri = 0.057107
 tripri = 0.12938
/

&dist_fn_knobs
 adiabatic_option ="iphi00=2"
 gridfac = 1.0
 boundary_option = "linked"
 mach = -0.079881
 g_exb = -0.0788355 ! corresponds to gamma_E
/

&theta_grid_knobs
 equilibrium_option = 'eik'
/

&theta_grid_eik_knobs
 itor = 1
 iflux = 0
 irho = 2
 ppl_eq = F
 gen_eq = F
 efit_eq = F
 local_eq = T
 eqfile = 'dskeq.cdf'
 equal_arc = T
 bishop = 4
 s_hat_input = 0.57383
 beta_prime_input = -0.052589
 delrho = 1.e-3
 isym = 0
 writelots = F
/

&kt_grids_knobs
 grid_option = 'box'
/

&kt_grids_box_parameters
 ! total number of ky's: naky = (ny-1)//3 + 1
 ! total number of kx's: nakx = 2*(nx-1)//3 + 1
 ny = 72                      ! i.e. kymax = 2.0
 nx = 144                     ! i.e. >= 1 twist-and-shift links for kymax
 y0 = 11.111                  ! i.e. dky = 1/y0 = 0.09
 jtwist = 4                   ! i.e. dkx = 2*pi*shat*dky/jtwist = 0.0811
 mixed_flowshear = .true.     ! turns on continuous-in-time algo for flow shear
/

&fields_knobs
 field_option ='implicit'
 force_maxwell_reinit = .false.
/

&le_grids_knobs
 ngauss = 5
 negrid = 16
 vcut = 2.5
/

&init_g_knobs
 chop_side = F
 phiinit = 1.e-3
 ! location to save/read restart file (overwritten when restarting)
 restart_file = "nc/run.nc"
 ginit_option = "noise"       ! FOR RESTARTS : set to "many"
 clean_init = .true.
 read_many = .true.
/

&knobs
 fphi = 1.0
 fapar = 0.0
 faperp = 0.0
 delt = 0.025
 nstep = 200000
 avail_cpu_time = 86400       ! 24hrs, adapt to available resources
 delt_option = "default"      ! FOR RESTARTS : set to "check_restart"
/

&nonlinear_terms_knobs
 nonlinear_mode = 'on'
 cfl = 0.25
/

&reinit_knobs
 delt_adj = 2.0
 delt_minimum = 1.e-4
 delt_cushion = 10000
/

&layouts_knobs
 ! consider layout = 'lxyes' for better performance
 layout = 'xyles'
 local_field_solve = F
/

&hyper_knobs
 hyper_option = 'visc_only'
 const_amp = .false.
 isotropic_shear = .false.
 D_hypervisc = 0.05
/

&gs2_diagnostics_knobs
 write_fluxes = .true.
 print_flux_line = T
 write_nl_flux = T
 print_line = F
 write_line = F
 write_omega = F
 write_final_fields = T
 write_g = F
 write_verr = T
 nwrite = 50
 navg = 50
 nsave = 3000
 omegatinst = 500.0
 save_for_restart = .true.
 omegatol = -1.0e-3
 save_many = .true.
/

!! ------------------------------------------------- !!
!                     End of input file                     !
!! ------------------------------------------------- !!


\end{lstlisting}

\section{Source code} \label{sec:source_code}

The version of the \tw{GS2} code used for this work is available at \url{https://bitbucket.org/gyrokinetics/gs2/branch/ndc_branch}, with the newest commit at the time of writing being \texttt{0abdcda}. The associated version of ``Makefiles'' necessary for compilation is available at \url{https://bitbucket.org/gyrokinetics/makefiles/branch/ndc_branch} under the commit \texttt{ba24979}, and the additional ``utils'' files required to run the code are available at \url{https://bitbucket.org/gyrokinetics/utils/branch/ndc_branch} under the commit \texttt{8e41f9a}.

\end{appendices}

\bibliographystyle{naturemag}
\bibliography{my_bib}

\end{document}